\documentclass[a4paper,11pt]{article}
\usepackage{amssymb}
\usepackage{amsfonts}
\usepackage{graphicx}
\usepackage{amsmath}

\DeclareFontFamily{U}{rsf}{}
\DeclareFontShape{U}{rsf}{m}{n}{
  <5> <6> rsfs5 <7> <8> <9> rsfs7 <10-> rsfs10}{}
\DeclareMathAlphabet\Scr{U}{rsf}{m}{n} \makeatletter
\@addtoreset{equation}{section} \makeatother

\def\be{\begin{equation}}
\def\ee{\end{equation}}
\def\ba{\begin{array}}
\def\ea{\end{array}}

\newcommand{\bea}{\begin{eqnarray}}
\newcommand{\eea}{\end{eqnarray}}

\begin{document}

\begin{titlepage}
 \thispagestyle{empty}
 \begin{flushright}
     \hfill{CERN-PH-TH/2012-212}\\
 \end{flushright}

 \vspace{50pt}

 \begin{center}
     { \huge{\bf      {On Symmetries of Extremal Black Holes\\ \vspace{7pt}with One and Two Centers}}}

     \vspace{70pt}

     {\Large {\bf Sergio Ferrara$^{a,b}$ and Alessio Marrani$^{a}$}}

     \vspace{30pt}

  {\it ${}^a$ Physics Department, Theory Unit, CERN,\\
     CH -1211, Geneva 23, Switzerland\\\texttt{sergio.ferrara@cern.ch}\\\texttt{alessio.marrani@cern.ch}}

     \vspace{20pt}

   {\it ${}^b$ INFN - Laboratori Nazionali di Frascati,\\
     Via Enrico Fermi 40, I-00044 Frascati, Italy}

     \vspace{15pt}

     \vspace{100pt}

     {ABSTRACT}
 \end{center}

 \vspace{10pt}
\noindent After a brief introduction to the \textit{Attractor Mechanism}, we review the appearance of groups \textit{of type }$E_{7}$ as
generalized electric-magnetic duality symmetries in locally supersymmetric
theories of gravity, with particular emphasis on the symplectic structure of
fluxes in the background of extremal black hole solutions, with one or two
centers. In the latter case, the role of an \textit{%
\textquotedblleft horizontal"} symmetry $SL_{h}\left( 2,\mathbb{R}\right) $
is elucidated by presenting a set of two-centered relations governing the
structure of two-centered invariant polynomials.
 \vfill
\begin{center}{\sl
Based on Lectures given by SF and AM at the School "Black Objects in Supergravity" (BOSS 2011),\\
INFN -  LNF, Rome, Italy, May 9-13 2011.\\
To appear in the Proceedings}
\end{center}

\end{titlepage}

\baselineskip 6 mm

\section{Introduction\label{Intro}}

The \textit{Attractor Mechanism} (AM) \cite{AM-Refs} governs the dynamics in
the scalar manifold of Maxwell-Einstein (super)gravity theories. It keeps
standing as a crucial fascinating key topic. Along the last years, a number
of papers have been devoted to the investigation of attractor configurations
of extremal black $p$-branes in diverse space-time dimensions; for some
lists of Refs., see \textit{e.g.} \cite{Revs}.

The AM is related to dynamical systems with fixed points, describing the
equilibrium state and the stability features of the system under
consideration\footnote{%
We recall that a point $x_{fix}$ where the phase velocity $v\left(
x_{fix}\right) $ vanishes is called a \textit{fixed} point, and it gives a
representation of the considered dynamical system in its equilibrium state,
\begin{equation*}
v\left( x_{fix}\right) =0.
\end{equation*}
The fixed point is said to be an \textit{attractor} of some motion $x\left(
t\right) $ if
\begin{equation*}
lim_{t\rightarrow \infty }x(t)=x_{fix}.
\end{equation*}%
}. When the AM holds, the particular property of the long-range behavior of
the dynamical flows in the considered (dissipative) system is the following:
in approaching the fixed points, properly named \textit{attractors}, the
orbits of the dynamical evolution lose all memory of their initial
conditions, but however the overall dynamics remains completely
deterministic.

The first example of AM in supersymmetric systems was discovered in the
theory of static, spherically symmetric, asymptotically flat extremal dyonic
black holes in $N{=2}$ Maxwell-Einstein supergravity in $d=4$ and $5$
space-time dimensions (see the first two Refs. of \cite{AM-Refs}). In the
following, we will briefly present some basic facts about the $d=4$ case.

The multiplet content of a completely general $N=2$, $d=4$ supergravity
theory is the following (see \textit{e.g.} \cite{N=2-Big}, and Refs.
therein):

\begin{enumerate}
\item the \textit{gravitational} multiplet
\begin{equation}
\left( V_{\mu }^{a},\psi ^{A},\psi _{A},A^{0}\right) ,  \label{g-mult}
\end{equation}
described by the \textit{Vielbein} one-form $V^{a}$ ($a=0,1,2,3$) (together
with the spin-connection one-form $\omega ^{ab}$), the $SU(2)$ doublet of
gravitino one-forms $\psi ^{A},\psi _{A}$ ($A=1,2$, with the upper and lower
indices respectively denoting right and left chirality, \textit{i.e.} $%
\gamma _{5}\psi _{A}=-\gamma _{5}\psi ^{A}$), and the graviphoton one-form $%
A^{0}$;

\item $n_{V}$ \textit{vector} supermultiplets
\begin{equation}
\left( A^{I},\lambda ^{iA},\overline{\lambda }_{A}^{\overline{i}%
},z^{i}\right) ,  \label{v-mult}
\end{equation}
each containing a gauge boson one-form $A^{I}$ ($I=1,...,n_{V}$), a doublet
of gauginos (zero-form spinors) $\lambda ^{iA},\overline{\lambda }_{A}^{%
\overline{i}}$, and a complex scalar field (zero-form) $z^{i}$ ($%
i=1,...,n_{V}$). The scalar fields $z^{i}$ can be regarded as coordinates on
a complex manifold $\mathcal{M}_{n_{V}}$ ($dim_{\mathbb{C}}\mathcal{M}%
_{n_{V}}=n_{V}$), which is actually a \textit{special K\"{a}hler} manifold;

\item $n_{H}$ \textit{hypermultiplets}
\begin{equation}
\left( \zeta _{\alpha },\zeta ^{\alpha },q^{u}\right) ,  \label{h-mult}
\end{equation}%
each formed by a doublet of zero-form spinors, that is the hyperinos $\zeta
_{\alpha },\zeta ^{\alpha }$ ($\alpha =1,...,2n_{H}$), and four real scalar
fields $q^{u}$ ($u=1,...,4n_{H}$), which can be considered as coordinates of
a quaternionic manifold $\mathcal{H}_{n_{H}}$ ($dim_{\mathbb{H}}\mathcal{H}%
_{n_{H}}=n_{H}$).
\end{enumerate}

\textit{At least} in absence of gauging and without quantum corrections, the
$n_{H}$ hypermultiplets are spectators in the AM. This can be understood by
looking at the transformation properties of the Fermi fields: the hyperinos $%
\zeta _{\alpha },\zeta ^{\alpha }$'s transform independently on the vector
fields, whereas the gauginos' supersymmetry transformations depend on the
Maxwell vector fields. Consequently, the contribution of the hypermultiplets
can be dynamically decoupled from the rest of the physical system; in
particular, it is also completely independent from the evolution dynamics of
the complex scalars $z^{i}$'s coming from the vector multiplets (\textit{i.e.%
} from the evolution flow in $\mathcal{M}_{n_{V}}$). Indeed, disregarding
for simplicity's sake the fermionic and gauging terms, the supersymmetry
transformations of gauginos and hyperinos respectively read (see \textit{e.g.%
} \cite{N=2-Big}, and Refs. therein)
\begin{eqnarray}
\delta \lambda ^{iA} &=&i\partial _{\mu }z^{i}\gamma ^{\mu }\varepsilon
^{A}+G_{\mu \nu }^{-i}\gamma ^{\mu \nu }\varepsilon _{B}\epsilon ^{AB};
\label{gauginos} \\
\delta \zeta _{\alpha } &=&i\mathcal{U}_{u}^{B\beta }\partial _{\mu
}q^{u}\gamma ^{\mu }\varepsilon ^{A}\epsilon _{AB}\mathbb{C}_{\alpha \beta }.
\label{hyperinos}
\end{eqnarray}%
(\ref{hyperinos}) implies that the asymptotical configurations of the
quaternionic hypermultiplets' scalars are \textit{unconstrained}, and
therefore they can vary continuously in the manifold $\mathcal{H}_{n_{H}}$
of the related quaternionic non-linear sigma model.

Thus, as far as ungauged theories are concerned, for the treatment of AM one
can restrict to consider $N=2$, $d=4$ Maxwell-Einstein supergravity, in
which $n_{V}$ vector multiplets (\ref{v-mult}) are coupled to the gravity
multiplet (\ref{g-mult}). The relevant dynamical system to be considered is
the one related to the radial evolution of the configurations of complex
scalar fields of such $n_{V}$ vector multiplets. When approaching the event
horizon of the black hole, the scalars dynamically run into fixed points,
taking values which are only function (of the ratios) of the electric and
magnetic charges associated to Abelian Maxwell vector potentials under
consideration.

The inverse distance to the event horizon is the fundamental evolution
parameter in the dynamics towards the fixed points represented by the
\textit{attractor }configurations of the scalar fields. Such near-horizon
configurations, which ``attracts'' the dynamical evolutive flows in $%
\mathcal{M}_{n_{V}}$, are completely independent on the initial data of such
an evolution, \textit{i.e.} on the spatial asymptotical configurations of
the scalars. Consequently, for what concerns the scalar dynamics, the system
completely loses memory of its initial data, because the dynamical evolution
is ``attracted'' by some fixed configuration points, purely depending on the
electric and magnetic charges.

In the framework of supergravity theories, extremal black holes can be
interpreted as BPS (Bogomol'ny-Prasad-Sommerfeld)-saturated \cite{BPS}
interpolating metric singularities in the low-energy effective limit of
higher-dimensional superstrings or $M$-theory \cite{GT-p-branes}. Their
asymptotically relevant parameters include the ADM mass \cite{ADM}, the
electrical and magnetic charges (defined by integrating the fluxes of
related field strengths over the $2$-sphere at infinity), and the
asymptotical values of the (dynamically relevant set of) scalar fields. The
AM implies that the class of black holes under consideration loses all its
``scalar hair'' within the near-horizon geometry. This means that the
extremal black hole solutions, in the near-horizon limit in which they
approach the Bertotti-Robinson $AdS_{2}\times S^{2}$ conformally flat metric
\cite{BR}, are characterized only by electric and magnetic charges, but not
by the continuously-varying asymptotical values of the scalar fields.

An important progress in the geometric interpretation of the AM was achieved
in the last Ref. of \cite{AM-Refs}, in which the attractor near-horizon
scalar configurations were related to the critical points of a suitably
defined black hole effective potential function $V_{BH}$. In general, $%
V_{BH} $ is a positive definite function of scalar fields and electric and
magnetic charges, and its non-degenerate critical points in $\mathcal{M}%
_{n_{V}}$
\begin{equation}
\forall i=1,...,n_{V},\frac{\partial V_{BH}}{\partial z^{i}}=0:~\left.
V_{BH}\right\vert _{\frac{\partial V_{BH}}{\partial z}=0}>0,
\label{crit-points}
\end{equation}%
fix the scalar fields to depend only on electric and magnetic fluxes
(charges). In the Einstein two-derivative approximation, the (semi)classical
Bekenstein-Hawking entropy ($S_{BH}$) - area ($A_{H}$) formula \cite{BH}
yields the (purely charge-dependent) black hole entropy $S_{BH}$ to be
\begin{equation}
S_{BH}=\pi \frac{A_{H}}{4}=\pi \left. V_{BH}\right\vert _{\frac{\partial
V_{BH}}{\partial z}=0}=\pi \sqrt{\left\vert \mathcal{I}_{4}\right\vert },
\label{BH-entropy-area-formula}
\end{equation}%
where $\mathcal{I}_{4}$ is the unique independent invariant homogeneous
polynomial (quartic in charges) in the relevant representation $\mathbf{R}$
of $G$ in which the charges sit (see Eq. (\ref{emb}) and discussion below).
The last step of (\ref{BH-entropy-area-formula}) does not apply to $d=4$
supergravity theories with quadratic charge polynomial invariant, namely to
the $N=2$ \textit{minimally coupled} sequence \cite{Luciani} and to the $N=3$
\cite{N=3} theory; in these cases, in (\ref{BH-entropy-area-formula}) $\sqrt{%
\left\vert \mathcal{I}_{4}\right\vert }$ gets replaced by $\left\vert
\mathcal{I}_{2}\right\vert $.

In presence of $n=n_{V}+1$ Abelian vector fields, the charge vector ($%
\Lambda =0,1,...,n_{V}$)%
\begin{equation}
Q\equiv \left( p^{\Lambda },q_{\Lambda }\right)  \label{Q}
\end{equation}
of magnetic ($p^{\Lambda }$) and electric ($q_{\Lambda }$) fluxes sits in a $%
2n$-dimensional representation $\mathbf{R}$ of the $U$-duality\footnote{%
Here $U$-duality is referred to as the \textquotedblleft
continuous\textquotedblright\ symmetries of \cite{CJ-1}. Their discrete
versions are the $U$-duality non-perturbative string theory symmetries
introduced by Hull and Townsend \cite{HT-1}.} group $G$, defining the
Gaillard-Zumino embedding \cite{GZ} of $G$ itself into $Sp\left( 2n,\mathbb{R%
}\right) $, which is the largest group acting linearly on the fluxes
themselves:
\begin{equation}
G\overset{\mathbf{R}}{\subsetneq }Sp\left( 2n,\mathbb{R}\right) .
\label{emb}
\end{equation}%
We consider here the (semi-)classical limit of large charges, also indicated
by the fact that we consider $Sp\left( 2n,\mathbb{R}\right) $, and not $%
Sp\left( 2n,\mathbb{Z}\right) $ (no Dirac-Schwinger-Zwanziger quantization
condition is implemented on the fluxes themselves).

After \cite{FG1,DFL,FM}, the the $\mathbf{R}$-representation space of the $U$%
-duality group is known to exhibit a \textit{stratification} into disjoint
classes of orbits, which can be defined through invariant sets of
constraints on the (lowest order, actually unique) $G$-invariant $\mathcal{I}
$ built out of the symplectic representation $\mathbf{R}$; this will be
reported in Sec. \ref{Orbits} It is here worth remarking the crucial
distinction between the \textquotedblleft large\textquotedblright\ orbits
and \textquotedblleft small\textquotedblright\ orbits. While the former have
$\mathcal{I}\neq 0$ and support an attractor behavior of the scalar flow in
the near-horizon geometry of the extremal black hole background\textit{\ }%
\cite{AM-Refs}, for the latter the Attractor Mechanism does not hold, they
have $\mathcal{I}=0$ and thus they correspond to solutions with vanishing
Bekenstein-Hawking \cite{BH} entropy (\textit{at least} at the Einsteinian
two-derivative level).\smallskip

\section{$U$-Duality and Groups of Type $E_{7}$}

From the treatment above, the black hole entropy $S_{BH}$ is invariant under
the electric-magnetic duality, in which the non-compact $U$-duality group
has a symplectic action both on the charge vector $Q$ (\ref{Q}) and on the
scalar fields (through the definition of a \textit{flat} symplectic bundle
\cite{Strominger-SKG} over the scalar manifold itself); see \textit{e.g.}
\cite{dW-review} for a review. The latter property makes relevant the
mathematical notion of groups of type $E_{7}$.

The first axiomatic characterization of groups of type $E_{7}$ through a
module (irrep.) was given in 1967 by Brown \cite{Brown-Groups-of-type-E7}. A
group $G$ of type $E_{7}$ is a Lie group endowed with a representation $%
\mathbf{R}$ such that:

\begin{enumerate}
\item $\mathbf{R}$ is \textit{symplectic}, \textit{i.e.} :
\begin{equation}
\exists !\mathbb{C}_{\left[ MN\right] }\equiv \mathbf{1\in R\times }_{a}%
\mathbf{R;}  \label{sympl-metric}
\end{equation}%
(the subscripts \textquotedblleft $s$\textquotedblright\ and
\textquotedblleft $a$\textquotedblright\ stand for symmetric and
skew-symmetric throughout) in turn, $\mathbb{C}_{\left[ MN\right] }$ defines
a non-degenerate skew-symmetric bilinear form (\textit{symplectic product});
given two different charge vectors $Q_{1}$ and $Q_{2}$ in $\mathbf{R}$, such
a bilinear form is defined as
\begin{equation}  \label{W-pre}
\left\langle Q_{1},Q_{2}\right\rangle \equiv Q_{1}^{M}Q_{2}^{N}\mathbb{C}%
_{MN}=-\left\langle Q_{2},Q_{1}\right\rangle ;
\end{equation}

\item $\mathbf{R}$ admits a unique rank-$4$ completely symmetric primitive $%
G $-invariant structure, usually named $K$-tensor
\begin{equation}
\exists !\mathbb{K}_{\left( MNPQ\right) }\equiv \mathbf{1\in }\left[ \mathbf{%
R\times R\times R\times R}\right] _{s}\mathbf{;}
\end{equation}
thus, by contracting the $K$-tensor with the same charge vector $Q$ in $%
\mathbf{R}$, one can construct a rank-4 homogeneous $G$-invariant
polynomial, named $\mathcal{I}_{4}$:
\begin{equation}
\mathcal{I}_{4}\left( Q\right) \equiv \frac{1}{2}\mathbb{K}%
_{MNPQ}Q^{M}Q^{N}Q^{P}Q^{Q},  \label{I4}
\end{equation}
which corresponds to the evaluation of the rank-$4$ symmetric form $\mathbf{q%
}$ induced by the $K$-tensor on four identical modules $\mathbf{R}$:
\begin{equation}
\mathcal{I}_{4}\left( Q\right) =\frac{1}{2}\left. \mathbf{q}\left(
Q_{1},Q_{2},Q_{3},Q_{4}\right) \right| _{Q_{1}=Q_{2}=Q_{3}=Q_{4}\equiv
Q}\equiv \frac{1}{2}\left[ \mathbb{K}%
_{MNPQ}Q_{1}^{M}Q_{2}^{N}Q_{3}^{P}Q_{4}^{Q}\right] _{Q_{1}=Q_{2}=Q_{3}=Q_{4}%
\equiv Q}.
\end{equation}
A famous example of \textit{quartic} invariant in $G=E_{7}$ is the \textit{%
Cartan-Cremmer-Julia} invariant \cite{Cartan}, constructed out of the
fundamental irrep. $\mathbf{R}=\mathbf{56}$.

\item if a trilinear map $T\mathbf{:R\times R\times R}\rightarrow \mathbf{R}
$ is defined such that
\begin{equation}
\left\langle T\left( Q_{1},Q_{2},Q_{3}\right) ,Q_{4}\right\rangle =\mathbf{q}%
\left( Q_{1},Q_{2},Q_{3},Q_{4}\right) ,
\end{equation}
then it holds that
\begin{equation}
\left\langle T\left( Q_{1},Q_{1},Q_{2}\right) ,T\left(
Q_{2},Q_{2},Q_{2}\right) \right\rangle =\left\langle
Q_{1},Q_{2}\right\rangle \mathbf{q}\left( Q_{1},Q_{2},Q_{2},Q_{2}\right) .
\end{equation}
This last property makes the group of type $E_{7}$ amenable to a treatment
in terms of (rank-3) Jordan algebras and related Freudenthal triple systems.
\end{enumerate}

Remarkably, groups of type $E_{7}$, appearing in $D=4$ supergravity as $U$%
-duality groups, admit a $D=5$ uplift to groups of type $E_{6}$, as well as
a $D=3$ downlift to groups of type $E_{8}$; see \cite{Truini}. It should
also be recalled that split form of exceptional Lie groups appear in the
exceptional Cremmer-Julia \cite{CJ} sequence $E_{D\left( D\right) }$ of $U$%
-duality groups of $M$-theory compactified on a $D$-dimensional torus, in $%
D=3,4,5$.

It is intriguing to notice that the first paper on groups of type $E_{7}$
was written about a decade before the discovery of of extended ($N=2$)
supergravity \cite{FVN}, in which electromagnetic duality symmetry was
observed \cite{FSZ}. The connection of groups of type $E_{7}$ to
supergravity can be summarized by stating that all $2\leq N\leq 8$-extended
supergravities in $D=4$ with symmetric scalar manifolds ${\frac{G}{H}}$ have
$G$ of type $E_{7}$ \cite{Borsten:2009zy,Ferrara:2011gv}, with the exception
of $N=2$ group $G=U(1,n)$ and $N=3$ group $G=U(3,n)$. These latter in fact
have a quadratic invariant Hermitian form $\left( Q_{1},\overline{Q}%
_{2}\right) $, whose imaginary part is the symplectic (skew-symmetric)
product and whose real part is the symmetric quadratic invariant $\mathcal{I}%
_{2}\left( Q\right) $ defined as follows
\begin{eqnarray}
\mathcal{I}_{2}\left( Q\right) &\equiv &\left[ \text{Re}\left( Q_{1},%
\overline{Q}_{2}\right) \right] _{Q_{1}=Q_{2}}; \\
\left\langle Q_{1},\overline{Q}_{2}\right\rangle &=&-\text{Im}\left( Q_{1},%
\overline{Q}_{2}\right) .
\end{eqnarray}
Thus, the fundamental representations of pseudo-unitary groups $U(p,n)$,
which have a Hermitian quadratic invariant form, do not strictly qualify for
groups of type $E_{7}$.

In theories with groups of type $E_{7}$, the Bekenstein-Hawking black hole
entropy is given by
\begin{equation}
S=\pi \sqrt{\left| \mathcal{I}_{4}\left( Q\right) \right| },  \label{S-I4}
\end{equation}
as it was proved for the case of $G=E_{7(7)}$ (corresponding to $N=8$
supergravity) in \cite{Kallosh:1996uy}. For $N=2$ group $G=U(1,n)$ and $N=3$
group $G=U(3,n)$ the analogue of (\ref{S-I4}) reads
\begin{equation}
S=\pi \left| \mathcal{I}_{2}\left( Q\right) \right| .
\end{equation}

For $3 < N\leq 8$ the following groups of type $E_{7}$ are relevant: $%
E_{7\left( 7\right) }$, $SO^*(12)$, $SU(1,5)$, $SL(2,\mathbb{R})\times {%
SO(6,n)}$; see Table 1. In $N=2$ cases of \textit{symmetric} vector
multiplets' scalar manifolds, there are $6$ groups of type $E_{7}$ \cite%
{Gunaydin:1983rk} : $E_{7\left( -25\right) }$, $SO^*(12)$, $SU(3,3)$, $Sp(6,%
\mathbb{R})$, $SL(2,\mathbb{R})$, and $SL(2,\mathbb{R})\times SO(2,n)$; see
Table 2. Here $n$ is the integer describing the number of matter (vector)
multiplets for $N=4,3,2$.

\begin{table}[t]
\begin{center}
\begin{tabular}{|c||c|c|}
\hline $N$& $
\begin{array}{c}
\\
$G$ \\
~
\end{array}
$ & $
\begin{array}{c}
\\
  \mathbf{ R}
\\
~
\end{array}

$ \\ \hline\hline $
\begin{array}{c}
\\
3 \\
~
\end{array}
$ & $U(3,n)$ & $ \mathbf{(3+n)}$     \\ \hline $
\begin{array}{c}
\\
4 \\
~
\end{array}
$ & $SL(2, \mathbb{R})\times {SO(6,n)}$ & $\mathbf{(2, 6+n)}$   \\
\hline $
\begin{array}{c}
\\
5 \\
~
\end{array}
$ & $SU(1,5)$ & $ \mathbf{ 20}$   \\ \hline $
\begin{array}{c}
\\
6 \\
~
\end{array}
$ & $SO^*(12)$ & $\mathbf{ 32}$
 \\ \hline
$
\begin{array}{c}
\\
8 \\
~
\end{array}
$ & $E_{7\left( 7\right) }$ & $\mathbf{ 56}$   \\ \hline
\end{tabular}
\end{center}
\caption{ $N\geq 3$ supergravity sequence of groups $G$ of  the
corresponding ${G\over H}$ symmetric spaces, and their symplectic
representations  $\mathbf{R}$}
\end{table}

\begin{table}[t]
\begin{center}
\begin{tabular}{|c||c|}
\hline $
\begin{array}{c}
\\
$G$\\
~
\end{array}
$ & $\mathbb{\mathbb{}}
\begin{array}{c}
\\
  \mathbf{ R}\\
~
\end{array}
$ \\ \hline\hline $
\begin{array}{c}
\\
{U(1,n)}
\end{array}
$ & $
\begin{array}{c}

\mathbf{(1+n)}\\

\end{array}
$ \\ \hline $
\begin{array}{c}
\\
{SL(2, \mathbb{R})}\times SO(2,n) ~ ~

\end{array}
$ & $
\begin{array}{c}
\mathbf{(2, 2+n)}

\end{array}
$ \\ \hline $
\begin{array}{c}
\\
SL(2, \mathbb{R}) ~
\end{array}
$ & $
\begin{array}{c}
\\
\mathbf{4}
\end{array}
$ \\ \hline $
\begin{array}{c}
\\
Sp(6,\mathbb{R})~
\end{array}
$ & $
\begin{array}{c}
\\
\mathbf{14}' \\

\end{array}
$ \\ \hline $
\begin{array}{c}
\\
SU(3,3)\end{array} $ & $
\begin{array}{c}
\\
\mathbf{20} ~
\end{array}
$ \\ \hline $
\begin{array}{c}
\\
SO^{\ast }(12)~
\end{array}
$ & $
\begin{array}{c}
\\
\mathbf{32} ~
\end{array}
$ \\ \hline $
\begin{array}{c}
\\
E_{7\left( -25\right) } ~
\end{array}
$ & $
\begin{array}{c}
\\
\mathbf{56} ~
\end{array}
$ \\ \hline
\end{tabular}
\end{center}
\caption{$N=2$ choices of groups $G$ of the  ${G\over H}$ symmetric
spaces and their symplectic representations  $\mathbf{R}$. The last
four lines refer to ``magic" $N=2$ supergravities.}
\end{table}

\section{\label{Orbits}Duality Orbits}

We here report some results on the stratification of the $\mathbf{R}$ irrep.
space of simple groups $G$ $E_{7}$. For a recent account, with a detailed
list of Refs., see \textit{e.g.} \cite{Small-Orbits-Phys}.

In supergravity, this corresponds to $U$-duality invariant constraints
defining the charge orbits of a single-centered extremal black hole, namely
of the $G$-invariant conditions defining the \textit{rank} of the dyonic
charge vector $Q$ (\ref{Q}) in $\mathbf{R}$ as an element of the
corresponding Freudenthal triple system (FTS) (see \cite{Ferrar,Krut}, and
Refs. therein). The symplectic indices $M=1,...,\mathbf{f}$ ($\mathbf{f}%
\equiv $dim$_{\mathbb{R}}\mathbf{R}\left( G\right) $) are raised and lowered
with the symplectic metric $\mathbb{C}_{MN}$ defined by (\ref{sympl-metric}%
). By recalling the definition (\ref{I4}) of the unique primitive rank-4 $G$%
-invariant polynomial constructed with $Q$ in $\mathbf{R}$, the \textit{rank}
of a non-null $Q$ as an element of FTS$\left( G\right) $ ranges from four to
one, and it is manifestly $G$-invariantly characterized as follows:

\begin{enumerate}
\item \textit{rank}$\left( Q\right) =4$. This corresponds to ``large''
extremal black holes, with non-vanishing area of the event horizon
(exhibiting \textit{Attractor Mechanism} \cite{AM-Refs}):
\begin{equation}
\mathcal{I}_{4}\left( Q\right) < 0, ~ ~ or ~ ~ \mathcal{I}_{4}\left(
Q\right) >0  \label{rank=4}
\end{equation}

\item \textit{rank}$\left( Q\right) =3$. This corresponds to ``small''
\textit{lightlike} extremal black holes, with vanishing area of the event
horizon:
\begin{equation}
\begin{array}{l}
\mathcal{I}_{4}\left( Q\right) =0; \\
\\
T\left( Q,Q,Q\right) \neq 0.%
\end{array}%
\end{equation}

\item \textit{rank}$\left( Q\right) =2$. This corresponds to ``small''
\textit{critical} extremal black holes:
\begin{equation}
\begin{array}{l}
T\left( Q,Q,Q\right) =0; \\
\\
3T\left( Q,Q,P\right) +\left\langle Q,P\right\rangle Q\neq 0.%
\end{array}%
\end{equation}

\item \textit{rank}$\left( Q\right) =1$. This corresponds to
\textquotedblleft small\textquotedblright\ \textit{doubly-critical} extremal
BHs \cite{FG1,FM}:
\begin{equation}
3T\left( Q,Q,P\right) +\left\langle Q,P\right\rangle Q=0,~\forall P\in
\mathbf{R}.  \label{rank=1}
\end{equation}
\end{enumerate}

Let us consider the doubly-criticality condition (\ref{rank=1}) more in
detail. \textit{At least} for \textit{simple} groups of type $E_{7}$, the
following holds:
\begin{eqnarray}
\mathbf{R}\times _{s}\mathbf{R} &=&\mathbf{Adj}+\mathbf{S};  \label{symm-1}
\\
\mathbf{R}\times _{a}\mathbf{R} &=&\mathbf{1}+\mathbf{A},  \label{skew-symm}
\end{eqnarray}%
where $\mathbf{S}$ and $\mathbf{A}$ are suitable irreps.. For example, for $%
G=E_{7}$ ($\mathbf{R}=\mathbf{56}$, $\mathbf{Adj}=\mathbf{133}$) one gets
\begin{eqnarray}
\left( \mathbf{56}\times \mathbf{56}\right) _{s} &=&\mathbf{133}+\mathbf{1463%
}; \\
\left( \mathbf{56}\times \mathbf{56}\right) _{a} &=&\mathbf{1}+\mathbf{1539}.
\end{eqnarray}%
For such groups, one can construct the projection operator on $\mathbf{Adj}%
\left( G\right) $:
\begin{eqnarray}
\mathcal{P}_{AB}^{~~CD} &=&\mathcal{P}_{\left( AB\right) }^{~~\left(
CD\right) }; \\
\mathcal{P}_{AB}^{~~CD}\frac{\partial ^{2}\mathcal{I}_{4}}{\partial
Q^{C}\partial Q^{D}} &=&\left. \frac{\partial ^{2}\mathcal{I}_{4}}{\partial
Q^{A}\partial Q^{B}}\right\vert _{\mathbf{Adj}\left( G\right) }; \\
\mathcal{P}_{AB}^{~~CD}\mathcal{P}_{CD}^{~~EF}\frac{\partial ^{2}\mathcal{I}%
_{4}}{\partial Q^{E}\partial Q^{F}} &=&\mathcal{P}_{AB}^{~~EF}\frac{\partial
^{2}\mathcal{I}_{4}}{\partial Q^{E}\partial Q^{F}},
\end{eqnarray}%
where (recall (\ref{symm-1}))
\begin{eqnarray}
\frac{\partial ^{2}\mathcal{I}_{4}}{\partial Q^{A}\partial Q^{B}} &=&\left.
\frac{\partial ^{2}\mathcal{I}_{4}}{\partial Q^{A}\partial Q^{B}}\right\vert
_{\mathbf{Adj}\left( G\right) }+\left. \frac{\partial ^{2}\mathcal{I}_{4}}{%
\partial Q^{A}\partial Q^{B}}\right\vert _{\mathbf{S}\left( G\right) }; \\
\left. \frac{\partial ^{2}\mathcal{I}_{4}}{\partial Q^{A}\partial Q^{B}}%
\right\vert _{\mathbf{Adj}\left( G\right) } &=&2\left( 1-\tau \right) \left(
3\mathbb{K}_{ABCD}+\mathbb{C}_{AC}\mathbb{C}_{BD}\right) Q^{C}Q^{D}; \\
\left. \frac{\partial ^{2}\mathcal{I}_{4}}{\partial Q^{A}\partial Q^{B}}%
\right\vert _{\mathbf{S}\left( G\right) } &=&2\left[ 3\tau \mathbb{K}%
_{ABCD}+\left( \tau -1\right) \mathbb{C}_{AC}\mathbb{C}_{BD}\right]
Q^{C}Q^{D},
\end{eqnarray}%
where $\tau \equiv 2\mathbf{d/}\left[ \mathbf{f}\left( \mathbf{f}+1\right) %
\right] $, $\mathbf{d}\equiv $dim$_{\mathbb{R}}\left( \mathbf{Adj}\left(
G\right) \right) $. The explicit expression of $\mathcal{P}_{AB}^{~~CD}$
reads\footnote{%
For related results in terms of a map formulated in the \textquotedblleft $%
4D/5D$ special coordinates\textquotedblright\ symplectic frame (and thus
manifestly covariant under the $d=5$ $U$-duality group $G_{5}$), see \textit{%
e.g.} \cite{Shukuzawa,Yokota}.} ($\alpha =1,...,\mathbf{d}$):
\begin{equation}
\mathcal{P}_{AB}^{~~CD}=\tau \left( 3\mathbb{C}^{CE}\mathbb{C}^{DF}\mathbb{K}%
_{EFAB}+\delta _{(A}^{C}\delta _{B)}^{D}\right) =-t^{\alpha \mid
CD}t_{\alpha \mid AB},  \label{P-Adj}
\end{equation}%
where the relation \cite{Exc-Reds} (see also \cite{Og-1})
\begin{equation}
\mathbb{K}_{MNPQ}=-\frac{1}{3\tau }t_{(MN}^{\alpha }t_{\alpha \mid PQ)}=-%
\frac{1}{3\tau }\left[ t_{MN}^{\alpha }t_{\alpha \mid PQ}-\tau \mathbb{C}%
_{M(P}\mathbb{C}_{Q)N}\right] ,  \label{rel-2}
\end{equation}%
where
\begin{equation}
t_{MN}^{\alpha }=t_{\left( MN\right) }^{\alpha };~~t_{MN}^{\alpha }\mathbb{C}%
^{MN}=0
\end{equation}%
is the symplectic representation of the generators of the Lie algebra $%
\mathfrak{g}$ of $G$. Notice that $\tau <1$ is nothing but the ratio of the
dimensions of the adjoint $\mathbf{Adj}$ and rank-$2$ symmetric $\mathbf{R}%
\times _{s}\mathbf{R}$ (\ref{symm-1}) reps. of $G$, or equivalently the
ratio of upper and lower indices of $t_{MN}^{\alpha }$'s themselves.

\section{\label{2-Center-Relations}From One to Two Centers}

In multi-centered black hole solutions \cite{Multi-Ctr}, a charge vector $%
Q_{a}$ can be associated to each center, with the index $a=1,..,p$, with $p$
denoting the number of centers. This index transforms in the fundamental
representation $\mathbf{p}$ of the so-called \textit{\textquotedblleft
horizontal"} symmetry $SL_{h}\left( p,\mathbb{R}\right) $ introduced in \cite%
{FMOSY-1} (see also \cite{Levay}).

We will here focus on the simplest case $p=2$, presenting a number of
fundamental relations defining the structure of electric-magnetic fluxes of
two-centered black hole solutions \cite{Small-1}.

From \cite{FMOSY-1,Irred-1}, we define the symmetric $\mathbf{I}_{abcd}$
tensor, sitting in the spin $s=2$ irrep. $\mathbf{5}$ of $SL_{h}(2,\mathbb{R}%
)$, as%
\begin{equation}
\mathbf{I}_{abcd}\equiv \frac{1}{2}\mathbb{K}%
_{MNPQ}Q_{a}^{M}Q_{b}^{N}Q_{c}^{P}Q_{d}^{Q}.  \label{I_abcd}
\end{equation}%
Thus, its first derivative reads
\begin{equation}
\widetilde{Q}_{M\mid abc}\equiv \frac{1}{4}\frac{\partial \mathbf{I}_{abcd}}{%
\partial Q_{d}^{M}}=\frac{1}{2}\mathbb{K}_{MNPQ}Q_{a}^{N}Q_{b}^{P}Q_{c}^{Q}=%
\widetilde{Q}_{M\mid \left( abc\right) },  \label{spin=3/2}
\end{equation}%
sitting in the spin $s=3/2$ irrep. $\mathbf{4}$ of $SL_{h}\left( 2,\mathbb{R}%
\right) $ (the horizontal indices $a=1,2$ are raised and lowered with $%
\epsilon ^{ab}$, with $\epsilon ^{12}\equiv 1$). For clarity's sake, we
report the explicit expressions of the various components of $\mathbf{I}%
_{abcd}$ (\ref{I_abcd}), as well as their relations with the components of $%
\widetilde{Q}_{abc}$ (\ref{spin=3/2}) (the subscripts \textquotedblleft $%
+2,+1,0,-1,-2$\textquotedblright\ denote the horizontal helicity of the
various components \cite{FMOSY-1,Irred-1}):%
\begin{equation}
\mathbf{I}_{+2}\equiv \mathcal{I}_{4}\left( Q_{1}\right) \equiv \mathbf{I}%
_{1111}=\left\langle \widetilde{Q}_{111},Q_{1}\right\rangle ;  \label{I+2}
\end{equation}%
\begin{equation}
\mathbf{I}_{+1}\equiv \mathbf{I}_{1112}=\left\langle \widetilde{Q}%
_{111},Q_{2}\right\rangle =\left\langle \widetilde{Q}_{112},Q_{1}\right%
\rangle ;  \label{I+1}
\end{equation}%
\begin{equation}
\mathbf{I}_{0}\equiv \mathbf{I}_{1122}=\left\langle \widetilde{Q}%
_{112},Q_{2}\right\rangle =\left\langle \widetilde{Q}_{122},Q_{1}\right%
\rangle ;  \label{I+0}
\end{equation}%
\begin{equation}
\mathbf{I}_{-1}\equiv \mathbf{I}_{1222}=\left\langle \widetilde{Q}%
_{122},Q_{2}\right\rangle =\left\langle \widetilde{Q}_{222},Q_{1}\right%
\rangle ;  \label{I-1}
\end{equation}

\begin{equation}
\mathbf{I}_{-2}\equiv \mathcal{I}_{4}\left( Q_{2}\right) \equiv \mathbf{I}%
_{2222}=\left\langle \widetilde{Q}_{222},Q_{2}\right\rangle .  \label{I-2}
\end{equation}

Thus, one can consider the following symplectic product of spin $3/2$
horizontal charge tensors:
\begin{equation}
\left\langle \widetilde{Q}_{abc},\widetilde{Q}_{def}\right\rangle \equiv
\widetilde{Q}_{M\mid abc}\widetilde{Q}_{N\mid def}\mathbb{C}^{MN}.
\end{equation}%
\textit{A priori}, $\left\langle \widetilde{Q}_{abc},\widetilde{Q}%
_{def}\right\rangle $ should project onto spin $s=3,2,1,0$ irreps. of $%
SL_{h}\left( 2,\mathbb{R}\right) $ itself; however, due to the complete
symmetry of the $K$-tensor (and to the results of \cite%
{Brown-Groups-of-type-E7,Exc-Reds}), the projections on spin $s=3$ and $1$
do vanish:
\begin{eqnarray}
s &=&3:\left\langle \widetilde{Q}_{(abc},\widetilde{Q}_{def)}\right\rangle
=0;  \label{s=3} \\
s &=&2:\left\langle \widetilde{Q}_{(ab\mid c},\widetilde{Q}_{d\mid
ef)}\right\rangle \epsilon ^{cd}=\frac{2}{3}\mathcal{W}\mathbf{I}_{abef};
\label{s=2} \\
s &=&1:\left\langle \widetilde{Q}_{(a\mid bc},\widetilde{Q}_{de\mid
f)}\right\rangle \epsilon ^{bd}\epsilon ^{ce}=0;  \label{s=1} \\
s &=&0:\left\langle \widetilde{Q}_{abc},\widetilde{Q}_{def}\right\rangle
\epsilon ^{ad}\epsilon ^{be}\epsilon ^{cf}=8\mathbf{I}_{6},  \label{s=0}
\end{eqnarray}%
where the symplectic product $\mathcal{W}$ and the sextic horizontal
polynomial $\mathbf{I}_{6}$ \cite{Irred-1} are respectively defined as (also
cfr. (\ref{W-pre}))%
\begin{equation}
\mathcal{W}\equiv \left\langle Q_{1},Q_{2}\right\rangle =\frac{1}{2}\mathbb{C%
}_{MN}\epsilon ^{ab}Q_{a}^{M}Q_{b}^{N};  \label{W}
\end{equation}%
\begin{equation}
\mathbf{I}_{6}\equiv \frac{1}{8}\left\langle \widetilde{Q}_{abc},\widetilde{Q%
}_{def}\right\rangle \epsilon ^{ad}\epsilon ^{be}\epsilon ^{cf}=\frac{1}{4}%
\left\langle \widetilde{Q}_{111},\widetilde{Q}_{222}\right\rangle +\frac{3}{4%
}\left\langle \widetilde{Q}_{122},\widetilde{Q}_{112}\right\rangle .
\label{rel-1}
\end{equation}%
The complementary relation to (\ref{rel-1}), namely $\frac{1}{4}\left\langle
\widetilde{Q}_{111},\widetilde{Q}_{222}\right\rangle -\frac{3}{4}%
\left\langle \widetilde{Q}_{122},\widetilde{Q}_{112}\right\rangle $
consistently turns out to be proportional (through $\mathcal{W}$) to the
zero helicity component of $\mathbf{I}_{abcd}$ ; indeed, by setting $\left(
a,b,e,f\right) =\left( 1,1,2,2\right) $ in (\ref{s=2}), one obtains:
\begin{equation}
\frac{1}{2}\mathbf{I}_{0}\mathcal{W}=\frac{1}{4}\left\langle \widetilde{Q}%
_{111},\widetilde{Q}_{222}\right\rangle -\frac{3}{4}\left\langle \widetilde{Q%
}_{122},\widetilde{Q}_{112}\right\rangle .  \label{s=2,sz=0}
\end{equation}

We conclude by pointing out some consequences of the \textit{rank} of a
charge vector, say $Q_{1}$, on the set of two-centered invariant polynomials
defined above \cite{Small-1}:
\begin{eqnarray}
\text{\textit{rank}}\left( Q_{1}\right) &=&3\Rightarrow \mathbf{I}_{+2}=0;
\label{rank(Q)=3} \\
&&  \notag \\
\text{\textit{rank}}\left( Q_{1}\right) &=&2\Rightarrow \widetilde{Q}%
_{111}=0\Rightarrow \left\{
\begin{array}{l}
\mathbf{I}_{+2}=\mathbf{I}_{+1}=0; \\
\\
\mathbf{I}_{6}=-\frac{1}{2}\mathbf{I}_{0}\mathcal{W};%
\end{array}%
\right.  \notag \\
&&  \label{rank(Q)=2} \\
\text{\textit{rank}}\left( Q_{1}\right) &=&1\Rightarrow \left\{
\begin{array}{l}
\mathbf{I}_{+2}=\mathbf{I}_{+1}=0; \\
\\
\mathbf{I}_{0}=-\frac{1}{6}\mathcal{W}^{2}; \\
\\
\mathbf{I}_{6}=-\frac{1}{2}\mathbf{I}_{0}\mathcal{W}=\frac{1}{12}\mathcal{W}%
^{3}.%
\end{array}%
\right.  \label{rank(Q)=1}
\end{eqnarray}

\section*{Acknowledgments}

The reported results were obtained in different collaborations with Laura Andrianopoli,
Leron Borsten, Anna Ceresole, Riccardo D'Auria, Mike Duff, G. W. Gibbons,
Murat G\"{u}naydin, Renata Kallosh, Emanuele Orazi, William Rubens, Raymond Stora, A.
Strominger, Mario Trigiante, and Armen Yeranyan, which we gratefully
acknowledge.

This work is supported by the ERC Advanced Grant no. 226455, \textit{%
\textquotedblleft Supersymmetry, Quantum Gravity and Gauge
Fields\textquotedblright } (\textit{SUPERFIELDS}).

\end{document}